\documentclass[pra,bibnotes,twocolumn]{revtex4}
\usepackage[dvips]{pstcol}
\usepackage{pst-node}
\usepackage{pst-plot}
\usepackage{pst-coil}
\usepackage{graphicx}
\begin{document}
\draft
\def\ds{\displaystyle}
\title{Discrete Parametric Oscillation and Nondiffracting Beams in a  Glauber-Fock Oscillator}
\author{ Z. Oztas, C. Yuce}
\address{Department of Physics, Anadolu University, Turkey }
\email{cyuce@anadolu.edu.tr}
\date{\today}
%\pacs{ 03.65.Vf, 11.30.Er, 03.65.Ta}
\begin{abstract}
We consider a Glauber-Fock oscillator and show that diffraction can be managed. We show how to design arrays of waveguides where light beams experience zero diffraction. We find an exact analytical family of nondiffracting localized solution. We predict discrete parametric oscillation in the Glauber-Fock oscillator.  
\end{abstract}
\maketitle

Diffraction is a fundamental process in physics. It is well known that a wave is diffracted if it passes through an opening or if there is an obstacle in its path. Diffraction leads to the broadening of an initial intensity profile in space. Propagation-invariant waves have attracted a considerable interest over the years.  The most familiar example of nondiffracting wave is that of a plane wave propagating in free space. Other nontrivial nondiffracting waves for various physical systems have been intensively investigated by many authors. Durnin introduced diffractionless (nonspreading) solution of the free-space scalar wave equation almost three decades ago \cite{temel}. His solution is in the form of a Bessel function in the transverse direction and the corresponding intensity profile remains invariant during propagation unlike other beams that spread during propagation. In the framework of quantum mechanics, only the plane wave had been known as a nondiffracting wave in 1-D before Berry and Balazs. They theoretically showed another nonspreading solution was available for the Schrodinger equation describing a free particle \cite{berry}. Their solution is unique in the sense that it self-accelerates although no external potential exists. The accelerating behavior is not consistent with the Ehrenfest theorem, which describes the motion of the center of mass of the wave packet. The reason why the Ehrenfest theorem doesn't work is because of the non-integrability of the Airy function. \\
The physics of photon propagation is discrete lattices is very rich and has been extensively studied by many authors. Nondiffracting waves in media such as waveguide or nonlinear materials are interesting to study. Diffraction in such an array of waveguides is governed by hopping light from site to site through optical tunneling. Discrete diffraction is different from continuous diffraction. As an example, if light is initially excited at only one waveguide of a 1-D periodic array of waveguide, then light spreads into two main lobes with several secondary peaks between them. The idea to control discrete diffraction has attracted a special attention. The case of optical field propagation in a linearly coupled, infinite array of one dimensional waveguides was considered in \cite{aith} and anomalous diffraction (negative discrete diffraction) and diffraction-free cases were theoretically discussed and experimentally realized. Discrete diffraction was shown to be controlled in size and sign by the input conditions and diffractionless beams and focusing of normally diverging beams were discussed in homogeneous waveguide arrays \cite{perts}. The self-collimation effect where the spatial width of a light beam does not change over hundreds of free-space diffraction lengths was realized in a macroscobic photonic lattice \cite{raki}. Periodic photonic structures where the strength of diffraction can be made normal, anomalous or zero in a very broad frequency range was introduced \cite{perts2}.\\
One special system where discrete diffraction can be studied is the semi-infinite and asymmetric Glauber Fock lattice that has recently been introduced into optics community \cite{glauber0,glauber3}. The system is composed of an array of evanescently coupled waveguides with a square-root distribution of the coupling between adjacent guides \cite{glauber0}. The first experimental realization with a direct observation of the classical analogue of Fock state displacements
was presented in \cite{glauber3}. The Glauber-Fock photonic lattice is interesting in the sense that every excited waveguide represents a Fock state and the system admits an exact analytical solution. In \cite{glauberek0}, the Ermakov-Lewis invariant is constructed for the Glauber-Fock oscillator with propagation distance dependent tunneling amplitude and refractive index gradients. Some interesting effects can be observed in a Glauber-Fock system in the presence of the refractive index gradient, which was shown to be accomplished by varying the waveguide writing velocity of the femtosecond laser \cite{glauberbloch}. For example both periodic collapses and revivals in a discrete Glauber-Fock oscillator were observed in the intensity evolution. It was shown that periodically changing tunneling amplitude leads to Bloch-like oscillations and dynamic delocalization depending on the oscillation frequency and strength of refractive index gradient \cite{glauber1}. It is interesting to observe the Bloch-like oscillations in spite of the fact that the evanescently coupled waveguide array has nonuniform coupling and semi-infinite. Dynamic localization and quantum self-imaging are other interesting effects that occur in periodical lattice. These two effects were theoretically discussed and shown to be possible in a Glauber-Fock oscillator \cite{glauberDL}. Glauber-Fock oscillator was shown to be engineered by the method of shortcuts to adiabaticity \cite{glauber2}. Recently, geometric phase for a Glauber-Fock oscillator lattice was measured \cite{refek2}. Quantum Rabi model based on light transport in two decoupled semi infinite binary tight binding photonic lattice with a square-root distribution of the coupling between neighboring sites was experimentally realized in \cite{refek1}. The standard Glauber-Fock oscillator is semi-infinite and all the experiments mentioned above were realized in an effectively semi infinite system (the dimension in the transverse direction is long enough). The effect of truncation in a finite Glauber-Fock oscillator was discussed in \cite{glauber4}. The Glauber-Fock oscillator was generalized to include the nonlinear interaction and the corresponding system was theoretically explored in \cite{glauber5}. In this paper, we consider Glauber-Fock oscillator with propagation distance dependent tunneling amplitude and refractive index gradient. The purpose of this paper to find a way to obtain difractionless propagating initial excitations. We will find a solution that is capable of maintaining its spatial form during propagation. In this way, engineered diffraction can be realized. Secondly, we study parametric oscillation in the system.

\section{Model}

We consider a semi-infinite
Glauber-Fock oscillator array consisting of evanescently coupled waveguides. The tunneling amplitude through
which particles are transferred from site to site increases with the square root of the site number $\ds{n}$. We suppose that tunneling amplitude and linearly increasing refractive index gradient are $z$-dependent, where $\ds{z}$ is the normalized propagation distance. The equation satisfied by the complex field amplitude at the $n$-th waveguide is given by \cite{glauberek0}
\begin{eqnarray}\label{ham23}
i\partial_zc_n+F~n~c_n+J (\sqrt{n+1}c_{n+1}+\sqrt{n}c_{n-1})=0
\end{eqnarray}
where $n=0,1,2,...$, $\ds{J=J(z)}$ is the z-dependent first order tunneling amplitude, $F=F(z)$
is the z-dependent refractive index gradient and $c_n$ is the field amplitude at the $n$-the waveguide. Note that $\ds{c_n(z)=0}$ for $\ds{n<0}$.  Therefore our system is semi-infinite and asymmetric.\\
To find the solution, we follow the method introduced in \cite{glauberek0}. Let us write the state vector as $\ds{| \psi  >=\sum_{n=0}^{\infty} c_n (z) |n>}$, where the Fock state $\ds{|n>}$ corresponds to situation when only the waveguide with number $n$ is excited \cite{glauber3}. Substituting this solution into the equation (\ref{ham23}) yields the Schrodinger equation $\ds{H\psi=i\frac{\partial\psi}{{\partial} z}}$ with $\hbar=1$. The corresponding Hamiltonian reads
\begin{equation}\label{revf241}
H=-\left(  F(z) \hat{n}+J(z) \left(\hat{a}+\hat{a}^{\dagger}\right) \right)
\end{equation}
where the bosonic creation and annihilation operators satisfy $\ds{\hat{a}^{\dagger}|n>=\sqrt{n+1}|n+1>}$ and $\ds{\hat{a}|n>=\sqrt{n}|n-1>}$, respectively and the number operator satisfies $\ds{\hat{n}|n>=n|n>}$. We can transform this Hamiltonian using $\ds{\hat{a}=\frac{q+ip}{\sqrt{2}}}$ and $\ds{\hat{a}^{\dagger}=\frac{q-ip}{\sqrt{2}}}$, where $\ds{q}$ and $\ds{p}$ normalized position and momentum operators, respectively. Then the Hamiltonian can be rewrittten in the following form
\begin{equation}\label{revf242}
H=-\left (\frac{p^2}{2m}+\frac{m}{2}\omega^2 q^2+\sqrt{2} Jq-\frac{F}{2}\right)
\end{equation}
where the $z$-dependent mass and frequency are defined by $\ds{m=1/F(z)}$ and $\ds{\omega^2=F(z)^2}$.\\
The exact analytical solution of this Hamiltonian is available if we change $\ds{z{\rightarrow}Z=-z}$. This is the Hamiltonian of a quantum harmonic oscillator with $Z$-dependent mass, frequency, and external driving force. Let us now obtain an exact analytical solution. We first transform the coordinate according to $\ds{q^{\prime}=\frac{q-q_c}{L}}$, where the $Z$-dependent function $\ds{q_c(Z)}$ describes translation and L(Z) is a $Z$-dependent dimensionless scale factor to be determined later. More precisely, the center of the wave packet moves according to $q_c(Z)$ and the width of the wave packet changes according to $L(Z)$. Under this coordinate transformation, the Z-derivative operator transforms as $\ds{\partial_Z\rightarrow\partial_Z-L^{-1} (\dot{L}  q^{\prime}+\dot{q_c } ) \partial_{q^{\prime}}}$, where dot denotes derivation with respect to $Z$. In the accelerating frame, we will seek
the solution of the form 
\begin{equation}\label{s378ts}
\psi_n(q^{\prime},Z)=\exp{\left(i\Lambda\right)}~ \frac{\phi_n({q^{\prime}},Z) }{\sqrt{L}}
\end{equation}
where the position dependent phase reads
$\ds{\Lambda(q^{\prime},Z)=m\left(\alpha
{q^{\prime}}+ \frac{\beta}{2}{q^{\prime}}^2+S\right)}$, $\alpha$, $\beta$ and $S$ are $Z$-dependent functions to be determined. Substitute this ansatz into the corresponding Schrodinger equation and demand that the resulting equation includes harmonic and linear potential terms. Therefore we choose $\ds{\alpha=L\dot{q_c}}$ , $\ds{\beta= L\dot{L}}$ and $\ds{\dot{S}+\frac{\dot{m}}{m}S=\frac{1}{2} \dot{q_c}^2-\frac{\omega^2}{2}
q_c^2-\frac{\sqrt{2}Jq_c }{m}+\frac{F}{2m}
}$. The resulting equation reads
\begin{equation}\label{revfckdsj}
-\frac{1}{2mL^2}\frac{\partial^2\phi }{\partial {q^{\prime}}^2} +(\frac{m}{2}\Omega^2 {q^{\prime}}^2 +U {q^{\prime}})\phi =i\frac{\partial\phi }{\partial Z}
\end{equation}
where $\ds{\Omega^2=L(\ddot{L}+\frac{\dot{m}}{m} \dot{L}+\omega^2 L) }$ and $\ds{U=mL(\ddot{q_c}+\frac{\dot{m}}{m} \dot{q_c}+\omega^2 q_c+\sqrt{2}\frac{J}{m})}$. We can now determine $L$ and $q_c$.
\begin{equation}\label{smn49vs}
\ddot{L}+\frac{\dot{m}}{m} \dot{L}+\omega^2 L=\frac{1}{m^2L^3}
\end{equation}
\begin{equation}\label{revf2ad43}
\ddot{q_c}+\frac{\dot{m}}{m} \dot{q_c}+\omega^2 q_c+\sqrt{2}\frac{J}{m}=0
\end{equation}
The former one (known as the Ermakov equation) is easy to solve for the initial condition $\ds{\dot{L}(Z=0)=0}$. It is given by $\ds{L(Z)=1}$. Therefore, the quadratic term in $\Lambda(q^{\prime},Z)$ disappears. The solution of the latter equation will be discussed below. \\
With the choices (\ref{smn49vs},\ref{revf2ad43}), the linear potential is eliminated from the equation (\ref{revfckdsj}) and the resulting equation for $\phi_n(q^{\prime})$ can be solved analytically. It is given by
\begin{equation}\label{swd47o2qs}
\phi_n(q^{\prime},Z)=N_n\exp{\left(i \int {\frac{E_n}{mL^2}}dZ-\frac{{q^{\prime}}^2}{2}\right)}   ~H_{n}(q^{\prime})
\end{equation}
where $\ds{E_n=(n+\frac{1}{2})}$ and $H_n$ are the Hermite polynomials and $N_n$ is the normalization constant. Transforming backwards yields the exact solution. We have analytically found the exact solution. Our solution and the one in \cite{glauberek0} are equivalent but our solution is advantageous since it is written in terms of the width and the center of mass of the wave packet. As we shall see below, the equation (\ref{revf2ad43}) enables us to see the dynamics of the system clearly.\\
Let us now write the exact solution
\begin{equation}\label{sxpcv27941}
\psi_n=N_n\exp{ \left(im \dot{q_c}
q+i \epsilon_n-\frac{(q-q_c)^2}{2} \right)}~H_n(q-q_c)
\end{equation}
where $\ds{\epsilon_n=mS+ \int \frac{E_n}{m} dZ-m\dot{q_c}q_c}$ and $Z=-z$. This exact solution allows us to study some interesting effects in the Glauber -Fock oscillator lattice. The parameter $\ds{q_c}$. plays a key role on the study of oscillation, nondiffraction and parametric oscillation.\\
Let us now discuss the equivalence between $\psi_n$ and $\ds{n}$. Suppose that $\ds{q_c(Z)=0}$. In this case, the eigenfunctions are orthogonal to each other, i.e., $\ds{<\psi_j(q,Z)|\psi_{j^{\prime}}(q,Z)>}=\delta_{j,j^{\prime}}$. Therefore, we stress that $\ds{|n>\equiv \psi_n (q^{\prime}=q,-Z)}$. We stress that $\ds{q_c(Z)=0}$ if $J=0$. Hence we conclude that it is impossible to get a nondiffracting wave in the Glauber Fock lattice if only a single site is initially excited.

\subsubsection{Oscillation}

\begin{figure}[t]\label{21}
\includegraphics[width=9cm]{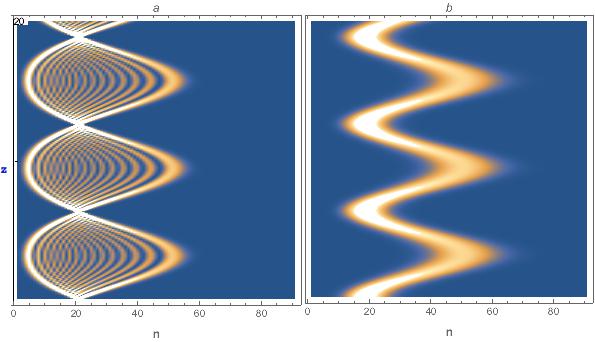}
\caption{ The density plots for the parameters $J=\sqrt{2}, f_0=1$. Periodical oscillation for the Glauber-Fock oscillator is shown in both figures. We assume $\ds{q_c(Z)=q_0( \cos{(\Omega_0 Z)}-\delta)}$ with $\delta=1$ (a) and $\delta=1/7$ (b). Therefore, both systems return to their initial states periodically at $z=2\pi,4\pi,...$. The system is initially excited at $20$-th lattice site in the figure (a). The initial wave packet in the fig.(b) is given by (\ref{sxpcv27941}) with $q_c(0)=6$. }
\end{figure}
Having established the exact analytical wave packet solution, we can now discuss solutions of the equation (\ref{revf2ad43}) in detail. That equation enables us to find the motion of the center of the wave packet.  Firstly, we study oscillations of the Glauber-Fock oscillator \cite{glauberbloch,glauber1}. If we suppose that the system is excited initially at a single site, then the initial conditions should be chosen as $\ds{q_c(0)=\dot{q_c}(0)=0}$. In this case, the initial wave function in the original frame is given by the equ. (\ref{sxpcv27941}) with $\ds{q_c(0)=\dot{q_c}(0)=0}$. Along $z$, the wave packet gains an extra $\ds{q}$-dependent phase. This term is responsible for diffraction. If $q_c(Z)$ is a periodic function, then the extra $\ds{q}$-dependent phase becomes equal to zero periodically. This implies that the wave packet returns to its initial form periodically. If we suppose either $q_c(0)\neq0$ or $\dot{q_c}(0)\neq0$ (but a periodic $q_c(z)$), then periodical oscillation occurs for the initial input that is a combination of excitation of many sites.\\
As an interesting example, let us choose $\ds{q_c(Z)=q_0( \cos{(\Omega_0 Z)}-\delta)}$, where the constant $\Omega_0$ is the oscillation frequency, the constant $q_0$ is the amplitude of the oscillation and the constant $\delta$ determines the initial value of $q_c(0)$. The case with $\delta=1$ ( $\ds{q_c(0)=\dot{q_c}(0)=0}$) refers to the system that is initially excited at a single site. Let us engineer our system to get this special $q_c(Z)$ function. Consider that $\ds{F=f_0}$, where $f_0$ is a constant. Then according to the equ. (\ref{revf2ad43}), the tunneling amplitude satisfies $\ds{\sqrt{2}J=q_0{\delta}f_0+q_0/f_0 (\Omega_0^2-f_0^2)\cos{(\Omega_0 Z)}} $. So, we conclude that either a constant $J$ ($\Omega_0=f_0$) or periodically modulated $J$ leads to the oscillation of the system. Let us confirm our prediction by numerically solving (\ref{ham23}) for $J=\sqrt{2}$ and $F=f_0=1$. We first consider single site initial excitation, which correspond to $\delta=\Omega_0=1$ and $q_0=2$. The fig-1.a plots the oscillatory behavior when only the $\ds{20}$-th waveguide is initially excited. As can be seen, the system periodically comes to its initial state (up to a phase factor) at regular intervals (at every $\ds{z=2\pi}$). We note that the oscillation is asymmetric because of the semi-infinite structure of the lattice. Suppose next that $\delta\neq1$. In this case, the system is not initially excited at a single site. Specifically, we choose $q_0=\delta^{-1}=7$ and $\Omega_0=f_0=1$. We numerically start with the initial wave function $\psi_{0}$ with $q_c(0)=6$. The Fig-1.b shows the propagation of this initial wave function. We see that the oscillatory dynamics of this broad wave is in full agreement with the above theoretical prediction.  \\
We have shown that discrete diffraction in the Glauber-Fock oscillator can be engineered. A proper choice of $\ds{q_c(Z)}$ allows us to analyze dynamics of the system. We can now engineer our system to predict some other interesting physical effect such as diffraction-free propagation and parametric oscillation. To do this, we should choose $\ds{q_c(Z)}$ appropriately. 

\subsubsection{Discrete Nondiffraction}

\begin{figure}[t]\label{20}
\includegraphics[width=6.5cm]{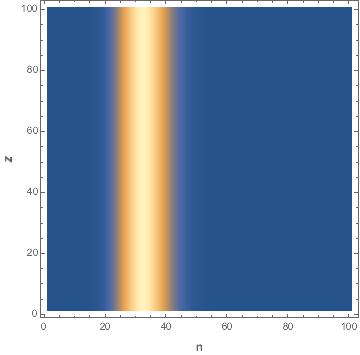}
\includegraphics[width=6.5cm]{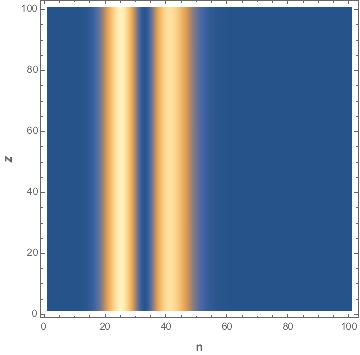}
\caption{ Both figures show the density $\ds{|\psi_n|^2}$ for the lowest order (top) and the first order (bottom) localized discrete nondiffracting beams as given by the equations (11) and (12), respectively. The first order nondiffracting beam has two separated beams. Similarly, the nondiffracting beam of order $n$ has (n-1) separated beams. The figures are plotted for $q_0=8$ ($J=2\sqrt{2}, f_0=0.5$). We emphasize that the width and the center of the nondiffracting beam is increased as $q_0$ is increased. As $q_0\rightarrow\infty$, the nondiffracting beam covers the whole lattice and becomes an extended nondiffracting wave. }
\end{figure}
Nondiffracting propagation can be studied if we assume that $\ds{q_c(Z)=q_0}$ is a constant. In this case, the phase term $\ds{\Lambda}$ is independent of $\ds{q}$ since $\dot{q}=0$. Furthermore, the density $\ds{|\psi_n|^2}$ is $Z$-independent. In other words, the wave function is propagation-invariant. The exact wave function for  this case is given by 
\begin{equation}\label{38qhd24}
\psi_n(q,z)=N_n\exp{\left(i\epsilon_n-\frac{(q-q_0)^2}{2}\right)}~H_n({q-q_0})
\end{equation} 
As can be seen, the initial and final wave packets are the same (up to a phase factor). Since the wave packet preserves its form at all $\ds{z}$, diffraction does not occur during propagation. This diffraction-free localized wave packet appears in infinitely long asymmetric system. \\
The equation (\ref{revf2ad43}) gives a constant solution $\ds{q_c(Z)=q_0=\frac{\sqrt{2}J}{F}}$ if $J=J_0$ and $F=f_0$ are constants (This is also true if $J/F$ is a constant). We note that there are infinitely many different diffraction-free waves. Changing $q_0$ generally increases the width of the nondiffracting wave and increasing the quantum number $n$ in $\psi_n$ allows us to obtain higher order nondiffracting waves. Let us visually see a nondiffracting propagation in our system. Consider the nondiffracting wave of order 0, i.e., $\ds{n=0}$. The initial wave packet reads $\ds{\psi_0(q,0)=N_0\exp{\left(-\frac{(q-q_0)^2}{2}\right)}}$. Let us expand this wave packet in terms of eigenkets $\ds{|n>}$
\begin{equation}\label{8dwf24d4}
\psi_0(q,0)=\sum_{n=0}  \frac{q_0^n}{(2^{n}n!)^{1/2}} e^{-\frac{q_0^2}{4}} |n>
\end{equation}
This last relation gives us information on how to excite our lattice for difractionless propagation. If we excite our system according to this formula, then nondiffracting propagation can be observed. One can see that diffractionless initial excitation gets more localized if $\ds{q_0}$ is decreased. The distance to the origin of the nondiffracting beam is also increased with the increase of $q_0$. We numerically evaluate the equation (\ref{ham23}) for the initial wave packet (\ref{8dwf24d4} )with $\ds{q_0=8}$ and $N=100$ sites. The top panel in the fig-2 depicts this nondiffracting propagation. As can be seen, the beam has the same intensity distribution along the propagation direction. \\
One may notice that the state (\ref{8dwf24d4}) is the standard coherent state. Therefore, the zeroth order non-diffracting state in our system is equivalent to the coherent state. But this is not the case for the higher order case. Consider the first excited state, $\ds{\psi_1(q,0)=N_1\exp{\left(-\frac{(q-q_0)^2}{2}\right)} H_1(q-q_0)  }$, which is the displaced Fock state. Expanding this wave packet in terms of eigenkets $\ds{|n>}$ yields
\begin{equation}\label{8dwf24scbj}
\psi_1(q,0)=\sum_{n=0}    \frac{-q_0^{n-1} ( q_0^2-2n)}{\sqrt{2^{n+1}  n!} \sqrt{1 + 2 q_0^2} }e^{-\frac{q_0^2}{4}} |n>
\end{equation}   
This state is not a coherent state since it does not satisfy $\ds{\hat{a}\psi_1(q,0)=const.~\psi_1(q,0)}$, where $\hat{a}$ is the annihilation operator. In the bottom panel of the fig.2, we plot the density for the first order nondiffracting wave with $q_0=8$. As can be seen there are two separated beams. Generally speaking, there are $n-1$ separated beams for the $n$-th order discrete nondiffracting beam.\\
We emphasize that discrete nondiffracting waves are not orthogonal to each other, $\ds{<\psi_j(q-q_0)|\psi_{j^{\prime}}(q-q_0)>\neq\delta_{jj^{\prime}}}$. Now, a question arises. Is a linear combination of two different discrete nondiffracting waves still nondiffracting? A combination of two nondiffracting waves with the same order but different $q_0$ is again a nondiffracting wave. However, a combination of two nondiffracting waves with different orders is not a nondiffracting wave any more.

\subsubsection{Discrete Parametric Oscillation}    
   
\begin{figure}[t]\label{320}
\includegraphics[width=9.5cm]{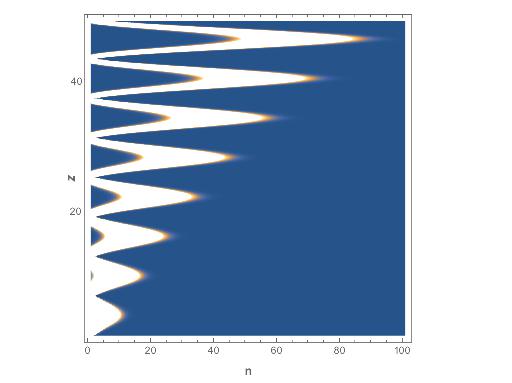}
\caption{ The density plot shows discrete parametric oscillation in the Glauber-Fock lattice when there are $N=100$ lattice sites. The system comes to its initial state periodically at every $2\pi$ along the propagation direction. The system undergoes oscillation with growing peaks during propagation. }
\end{figure}  
The equation (\ref{revf2ad43})  is a classical harmonic oscillator equation with time dependent mass and frequency. Therefore a question arises. Can we observe classical parametric oscillation in a discrete quantum system? The answer is yes for our system. A parametric oscillation is an oscillation whose amplitude grows during oscillation. This oscillation is different from Bloch oscillation since the amplitude of the Bloch oscillation does not grow during propagation. If the system is infinitely long, the amplitude of the oscillation goes to infinity as $z$ goes to infinity. This doesn't mean delocalization since the system comes to its initial state after each cycle. If the system is finite, parametric oscillation can still be observed until the amplitude of the oscillation reaches the length of finite lattice.\\
The condition for the oscillation in our system is that $\ds{q_c(0)=q_c(jT)}$ and $\ds{\dot{q}_c(0)=\dot{q}_c(jT)}$, where $T$ is the period and $j$ is a positive integer. This oscillation is a parametric oscillation if the amplitude of $\ds{q_c(Z)}$ grows linearly with $Z$ during propagation. As an example, consider one of the simplest choice: $\ds{q_c(Z)=q_0 ~Z(\cos(\Omega_0 Z)-\delta)}$, where $q_0$, $\Omega_0$ and $\delta$ are constants. As can be seen, the wave packet (\ref{sxpcv27941}) with this $q_c(Z)$ comes to its initial state periodically at every $2\pi/\Omega_0$. What is striking here is that the amplitude of the oscillation grows linearly with the propagation distance. In this way, the wave packet undergoes parametric oscillation. As discussed in the preceding subsection, $\delta=1$ corresponds to a single site initial excitation and and $\delta\neq1$ corresponds to a broad initial wave packet, respectively. This means that parametric oscillation can be observed in both cases. \\
Let us now see parametric oscillation graphically for the lowest order case. To observe discrete parametric oscillation, we need to choose $F$ and $J$ appropriately. Let us choose $\ds{q_c(t)=-\sqrt{2}J_0 + \epsilon Z   (\cos(Z) - 1)}$ and $F=1$, where $J_0$ and $\epsilon<<1$ are constants. According to the equation (\ref{revf2ad43}), the tunneling amplitude for the observation of discrete parametric oscillation is given by $\ds{ J=J_0 +\frac{ \epsilon Z + 2 \epsilon \sin(Z) }{\sqrt{2}}  }$, where $Z=-z$. To check our findings, we perform numerical computation for the parameters $J_0=-1/\sqrt{2}$ and $\epsilon=0.1$. The figure-3 plots our numerical simulation. One can see discrete parametric oscillation in the figure. As expected, the amplitude of the oscillation increases steadily during propagation.  The system comes to its initial state at every $2 \pi$ in the propagation direction.\\
To sum up, we have considered a Glauber Fock lattice and designed it to observe a couple of interesting effects such as nondiffraction and parametric oscillation. We have found the condition for the occurrence of discrete nondiffracting waves. The nondiffracting waves we study here are localized wave packets. We have also predicted discrete parametric oscillation in our system. Our findings can be verified experimentally with current technology. Here, we have analytically investigated Glauber-Fock oscillator. Parametric oscillation in other discrete systems is also worth studying.


\begin{thebibliography}{0}
\bibitem{temel} J. Durnin, J. Opt. Soc. Am. A \textbf{4} 651 (1987).
\bibitem{berry} M. V. Berry and N. L. Balazs,  Am. J. Phys. \textbf{47} 264 (1979).
\bibitem{aith} H. S. Eisenberg, Y. Silberberg, R. Morandotti, and J. S. Aitchison, Phys. Rev. Lett. \textbf{85} 1863 (2000).
\bibitem{perts} T. Pertsch et al., Phys. Rev. Lett. \textbf{88} 093901 (2002). 
\bibitem{raki}  P. T. Rakich et al., Nature Materials \textbf{5} 93 (2006).
\bibitem{perts2} I. L. Garanovich, A. A. Sukhorukov, and Yu. S. Kivshar, Phys. Rev. E \textbf{74} 066609 (2006). 
\bibitem{glauber0}  A. Perez-Leija, H. Moya-Cessa, A. Szameit, and D.N. Christodoulides, Opt. Lett. \textbf{35} 2409 (2010).
\bibitem{glauber3} R. Keil, A. Perez-Leija, F. Dreisow, M. Heinrich, H. Moya-Cessa, S. Nolte, D. N. Christodoulides, and A. Szameit, Phys. Rev. Lett. \textbf{107} 103601 (2011).
\bibitem{glauberek0} B. M. Rodriguez Lara, P. Aleahmad, H. M. Moya-Cessa, D. N. Christodoulides, Opt. Lett. \textbf{39} 2083 (2014).
\bibitem{glauberbloch} Robert Keil, Armando Perez-Leija, Parinaz Aleahmad, Hector Moya-Cessa, Stefan Nolte, Demetrios N. Christodoulides, and Alexander Szameit, Opt. Lett. \textbf{37} 3801 (2012).
\bibitem{glauber1} Armando Perez-Leija, Robert Keil, Alexander Szameit, Ayman F. Abouraddy, Hector Moya-Cessa, and Demetrios N. Christodoulides, Phys. Rev. A \textbf{85} 013848 (2012).
\bibitem{glauberDL} S Longhi and A Szameit, J. Phys.: Condens. Matter \textbf{25} 035603 (2013).
\bibitem{glauber2} Dionisis Stefanatos, Phys. Rev. A \textbf{90} 023811 (2014).
\bibitem{refek2} Kai Wang, Steffen Weimann, Stefan Nolte, Armando Perez-Leija, and Alexander Szameit, Opt. Lett. \textbf{41} 1889 (2016).
\bibitem{refek1} A. Crespi, S. Longhi, and R. Osellame, Phys. Rev. Lett. \textbf{108} 163601 (2012).
\bibitem{glauber4} B. M. Rodriguez-Lara, Phys. Rev. A \textbf{84} 053845 (2011).
\bibitem{glauber5} Alejandro J. Martinez, Uta Naether, Alexander Szameit, Rodrigo A. Vicencio, Opp. Lett. \textbf{37} 1865 (2012).
\end{thebibliography}
\end{document}